\documentclass[12pt, letterpaper, ]{article}

\usepackage[utf8]{inputenc}
\usepackage{amssymb}
\usepackage{multicol}
\usepackage[margin=1.8cm]{geometry}
\usepackage{setspace}
\usepackage{graphicx}
\usepackage{url}
\usepackage{color}
\usepackage{hyperref}
\graphicspath{ {images/} }
\usepackage{tikz}
\usepackage{amsmath}
\usepackage{physics}
\usepackage{ulem}

\hypersetup{colorlinks=true, linkcolor=black}

\begin{document}

\title{Is Bohr's Correspondence Principle just \\ Hankel's Principle of Permanence?}

\author{Iulian D. Toader \\ Institute Vienna Circle}

\date{}

\maketitle

\begin{abstract}

\noindent No, but the paper argues that Bohr understood his correspondence principle, or at least an aspect of that principle expressed by the notion of rational generalization, as grounded in Hankel's principle of permanence, adapted to new historical and theoretical contexts. This is shown to illuminate some otherwise obscure aspects of Bohr's approach to quantum theory, as well as a seemingly strange criticism against this approach, due to Feyerabend and Bohm.

\end{abstract}


\doublespacing

\section*{Introduction}

My main goal is to analyze an application of Hankel's \textit{Principle of Permanence} (HP) to quantum theory, an application that unfolded, I argue, under the guise of Bohr's more widely discussed \textit{Correspondence Principle} (CP). I start by presenting HP in the context of 19th Century mathematics, as a methodological principle, stipulating the preservation of rules as far as possible, but also as a metatheoretical principle, concerning the interpretability relations between theories. Then I briefly recall Bohr's uses of CP, emphasizing some changes that occurred over time, from its emergence in the old quantum theory to its application in the articulation and defense of his approach to quantum mechanics. After adducing some evidence that Bohr saw the shift from classical to quantum physics in the way Hankel, and Peacock before him, had understood the transition between mathematical theories, I argue that this illuminates not only Bohr's understanding of CP, but also his approach to quantum mechanics. More specifically, it sheds light on his otherwise obscure remarks on quantum mechanics as a rational generalization of classical physics, and clarifies the sense in which he thought the rules of quantum mechanics determined its physical meaning. On the background of this reading of CP as grounded in HP, I explain why Feyerabend's and Bohm's criticism of Bohr's doctrine of the necessity of classical concepts was well justified, and then I show how Howard's more recent reconstruction of this doctrine can be backed without any tension with CP.

\section{The Principle of Permanence in Historical Context}

The precursor of HP was a principle first explicitly formulated by George Peacock in 1833, which he called the \textit{Principle of the Permanence of Equivalent Forms} (PF):

\begin{quote} \singlespacing

Whatever equivalent form is discoverable in arithmetical algebra considered as the science of suggestion, when their symbols are general in their form, though specific in their value, will continue to be an equivalent form when the symbols are general in their nature as well as in their form. (Peacock 1833, 198sq, emphasis removed)

\end{quote}

\noindent Arithmetical algebra, or what had been called ``specious or universal arithmetic'', is a theory the language of which includes not only constants and signs for operations, just like ``elementary'' or ``common arithmetic'' does, but also variables ranging over the domain of positive integers. This contrasted with Peacock's symbolic algebra, the language of which further includes variables that are allowed to range over any domain of objects whatsoever, but most importantly over negative, rational, and imaginary numbers such as the ``impossible'' roots of equations of second or higher degree. As Peacock conceived of it, PF stipulates that some equivalent forms, i.e., at least the ones that can be discovered, which are expressed in the language of arithmetical algebra, should be preserved as equivalent forms, when expressed in the language of symbolic algebra. A prime example of such forms are the ``laws of combination'', what Peacock indiscriminately also called the principles or the rules of a theory. For instance, the arithmetically algebraic rule of distributivity, which he wrote as $ma+na=(m+n)a$, will be preserved in symbolic algebra as such. Peacock seems to have intended the notion of equivalent forms to be more general, so what is to be preserved in passing from one theory to another are not only such rules, but theorems as well. In any case, he clearly rejected the universal validity of PF, for he saw that there are equivalent forms of arithmetical algebra that are ``essentially arithmetical'' and these cannot be preserved when passing to symbolic algebra. Whenever PF could be applied, however, it was so useful for solving problems and proving theorems that other mathematicians, such as De Morgan, came to regard it not merely as a heuristic principle, but as a necessary mathematical truth.\footnote{For De Morgan's 1837 formulation of the principle of permanence, and for his discussion of this with Ada Lovelace, who questioned the validity of the principle and his insistence on ``the necessity of its truth'', see Hollings \textit{et al.} 2017. For discussion of criticisms raised against the principle of permanence by Peano and Hahn, see Toader 2021, 2023.} 

The sense in which arithmetical algebra is to be considered as a ``science of suggestion'' seems, nevertheless, rather obscure, so an explanation is needed. Peacock's concern was the clarification of the nature of the principles of symbolic algebra, which he thought had been misunderstood as being deduced from arithmetical principles -- a misunderstanding that he took to be responsible for the view of symbolic algebra as a generalization of arithmetical algebra and, thus, a generalization of elementary arithmetic. PF, and in particular the notion of a science of suggestion, was intended to help with Peacock's refutation of this view. On his own view, the principles of symbolic algebra are suggested by, rather than deduced from, the principles of arithmetical algebra and, as a consequence, symbolic algebra is something else than a generalization of arithmetical algebra. To say otherwise was, to him, an abuse of terminology, for the following reason: 

\begin{quote} \singlespacing
    
The operations in arithmetical algebra can be previously defined, whilst those in symbolic algebra, though bearing the same name, cannot: their meaning, however, when the nature of the symbols is known, can be generally, but by no means necessarily, interpreted. The process, therefore, by which we pass from one science to the other is not an ascent from particulars to generals, which is properly called \textit{generalization}, but one which is essentially arbitrary, though restricted with a specific view to its operations and their results admitting of such interpretations as may make its applications most generally useful. (\textit{op. cit.}, 194) 
\end{quote}

\noindent Peacock dismissed the view according to which one advances from arithmetical algebra to symbolic algebra by generalization, for this would entail that the operations of symbolic algebra (and their results) must admit of an interpretation over the domain of arithmetical algebra. In contrast, on his view, one rather passes from arithmetical algebra to symbolic algebra by suggestion, which does not entail that those operations and their results must admit of such an interpretation. Indeed, he considered any particular interpretation, arithmetical or non-arithmetical, as in principle dispensable. But Peacock qualified this view in an important way: arithmetical algebra, he noted, ``necessarily \textit{suggests} its principles'', i.e., the principles of symbolic algebra. However, this does not mean that arithmetical algebra is indispensable as a science of suggestion, for Peacock considered it simply the ``most convenient'', though in principle not the only possible science of suggestion for symbolic algebra.\footnote{See, e.g., Gregory 1840, which takes geometry as a science of suggestion for symbolic algebra. For more on Peacock and the history of British algebra, see e.g. Pycior 1981 and Lambert 2013.} What he meant, I think, is that the \textit{relation} of suggestion is necessary. 

On Peacock's view, then, any ``arbitrary assumptions'' that are consistent with the principles of arithmetical algebra, or as he put it, ``as far as they can exist in common'', might be stipulated as principles of symbolic algebra. But since any useful application of symbolic algebraic operations and results requires their interpretation over the domain of the application, the arbitrary assumptions must be restricted such that symbolic operations and results can admit of such an interpretation. This restriction is satisfied by necessarily assuming some subordinate science of suggestion, like arithmetical algebra. Thus, if the principles of symbolic algebra are not entirely arbitrary, but they are both consistent with and suggested by the principles of arithmetical algebra, then the operations of symbolic algebra and their results can be interpreted over the positive integers. 

Peacock's view, as I understand it, was something like this: the principles  of symbolic algebra are such that although it is not the case that symbolic algebraic operations must admit of an arithmetical interpretation, it must be the case that they can admit of an arithmetical interpretation. This is why one takes PF as the ``proper guide'' for the development of symbolic algebra, and arithmetical algebra as a science of suggestion, in the first place. \textit{The arithmetical interpretation is not necessary; it is only necessarily possible.} This is dictated by the requirement of usefulness in applications. Furthermore, if one takes PF as a proper guide for the development of symbolic algebra, then the latter's principles, albeit not deducible from the principles of arithmetical algebra, are taken by Peacock to be deducible from the conjunction of PF and the principles of arithmetical algebra. In this sense, PF is also acknowledged as the ``real foundation'' of symbolic algebra, the principles of which are considered, on the one hand, as arbitrary assumptions, and on the other hand, as ``necessary consequences'' of PF.

An important question about Peacock's view, as I have just described it, is the question about the meaning of symbolic algebra. If, as he maintained in the quotation above, its operations cannot be ``previously defined'', i.e., if their meaning cannot be given before the principles or the rules of symbolic algebra are stipulated, then what is it that gives meaning to these operations? Peacock's answer is as follows:

\begin{quote} \singlespacing
    
In arithmetical algebra, the definitions of the operations determine the rules; in symbolic algebra, the rules determine the meaning of the operations, or more properly speaking, they furnish the means of interpreting them  (\textit{op. cit.}, 200).
    
\end{quote}

\noindent This answer signifies a metasemantic revolution in 19th century mathematics, insofar as it represents a radical change from a traditional view, according to which meaning is determined by definitions relative to a particular domain, to a novel view according to which meaning is determined by rules. One way to understand Peacock's claim that the rules of symbolic algebra determine the meaning of its operations, in the sense that these rules ``furnish the means of interpreting'' these operations, is by taking the rules to determine the interpretation of operations. But if arithmetical algebra is considered as a science of suggestion, and if one understands this notion as I proposed above, then an interpretation over the positive integers of the operations of symbolic algebra is necessarily possible, without being necessary. Thus, to say that the rules of symbolic algebra determine the meaning of its operations is to say that the rules determine the interpretability of operations, or more precisely, the interpretability of the results obtained by these operations. Interpretability is necessary, as already noted, because without it, there would be no useful applications of symbolical algebra.

Peacock's view was famously endorsed by William Whewell:

\begin{quote} \singlespacing
    
\textit{The absolute universality of the interpretation of symbols} is the fundamental principle of their use. This has been shown very ably by Professor Peacock in his \textit{Algebra}. He has there illustrated, in a variety of ways, this principle: that `If general symbols express an identity when they are supposed to be of any special nature, they must also express an identity when they are general in their nature.' And thus this universality of symbols is a principle ... of the greatest importance in the formation of mathematical science, according to the wide generality which such science has in modern times assumed. (Whewell 1840, 143) 

\end{quote}

\noindent The universality of symbols, i.e., their ranging as variables over any domain of objects whatsoever, requires that the results of symbolic operations be interpretable over some particular domain, like that of arithmetical algebra. According to Peacock, as we have seen, this interpretability was indispensable for the applications of symbolic algebraic results, and enough to render the symbolic algebraic operations meaningful. Whewell's comments reinforced Peacock's view that PF is not only a methodological or heuristic principle, a useful guide or strategy for the development of new mathematical theories, but also a metatheoretical principle that properly characterizes the interpretability relations between theories.

Hankel adopted Peacock's view of the PF when he set up to develop the formal theory of complex numbers. He formulated his own version of the principle (HP) -- ``das hodegetische Princip der Permanenz der formalen Gesetze'' (Hankel 1867, vii) -- as follows: 

\begin{quote}
\singlespacing

If two forms expressed in general signs of the universal arithmetic are equal to one another, they should remain equal if the signs cease to denote simple quantities and the operations thereby receive some different content as well. (\textit{op. cit.}, 11)

\end{quote}

\noindent This corresponds, roughly, to Peacock's own formulation of the PF. For example, the rule of (left and right) distributivity in universal arithmetic, $a(b+c)=ab+ac$ and $(b+c)a=ba+ca$, where the quantities denoted by its variables are the positive integers, should be preserved when they denote any other objects, like the negative, rational, or imaginary numbers. Just what rules (or formal laws) are exactly to be preserved when extending the number domain is a question that Hankel carefully considered. Importantly, like Peacock, Hankel also warned against the universal application of HP, emphasizing that certain rules that hold for the real numbers cannot be extended to complex and hypercomplex numbers.\footnote{Cf. Hankel 1867, 195. He also proved that there can exist no extension beyond the complexes that preserves the commutativity of basic operations (Detlefsen 2005, 286). For recent discussions of Hankel's philosophy of mathematics, especially in connection to Frege's criticism of it, see Tappenden 2019 and Lawrence 2021.} This justified his initial characterization of HP as a methodological or heuristic (\textit{hodegetisches}) principle. But as we will presently see, Hankel, again like Peacock, regarded HP also as a metatheoretical principle that characterizes the relations between theories. Furthermore, Hankel came to think of it as a ``metaphysical'' principle, a view that was suggested to him by his envisaged application of HP to the physics of mechanical quantities.

To make sense of HP, and then identify in Hankel's writings the novel metasemantics noted above in Peacock's, we will need an account of his view of numbers, especially imaginary numbers, the possibility of which Hankel understood in terms of logical consistency: numbers exist only if their concept is clearly and distinctly defined without any contradiction. The question about the existence of numbers, he suggested, reduces to the question about the existence of the thinking subject or the objects of thought, since numbers represent the relations between such objects. He then distinguished, in Kantian terms, between two main types of numbers: on the one hand, what he called ``transcendent, purely mental, purely intellectual or purely formal'' numbers are those representing relations between objects of thought that cannot be constructed in intuition. On the other hand, actual numbers, or what Hankel called ``actuelle Zahlen'', are those representing relations between objects of thought that are constructed in intuition. However, he considered this distinction to be ``not a rigid, but a blurred distinction'' (``kein starrer, sondern ein fliessender'', \textit{op. cit.}, 8). Indeed, he further characterized as ``potentielle Zahlen'' those numbers that, although initially taken as purely formal, eventually become actual numbers, just as the complexes did after receiving a geometrical representation. Potential numbers formally represent relations between objects of thought, which are such that an intuitive construction of them turns out to be nevertheless possible. 

With his classification of types of numbers in place, Hankel characterized formal mathematics as a pure doctrine of forms (\textit{reine Formenlehre}): 

\begin{quote} \singlespacing
 
The condition for the establishment of a general arithmetic is therefore a purely intellectual mathematics, detached from all intuition, a pure doctrine of forms, in which it is not quantities or their representations [\textit{Bilder}], the numbers, that are combined, but intellectual objects, objects of thought, to
which actual [\textit{actuelle}] objects or relations thereof can, but do not have to, correspond. (\textit{op. cit.}, 10) \end{quote}

\noindent Thus, Hankel's formal mathematics stipulates rules of combination for potential numbers, i.e., for numbers representing relations an intuitive construction of which is possible, though not necessary. These rules are arbitrary to an extent limited solely by logical consistency, which Hankel thought could be established by their mutual independence. Nothing like Peacock's relation of suggestion seems to be explicitly required to impose further restrictions on the rules of formal mathematics. Nevertheless, Hankel believed that a system of operations obeying formal rules remains ``empty'', if no applications of results are possible. An empty system allows no relations between objects of thought to be constructed in intuition. Thus, Hankel understood formal mathematics as a theory of potential numbers, conceived of as purely formal representations of relations, for which it must be the case that a construction in intuition -- an interpretation -- of its formal results can be given. While he also thought that no such particular interpretation was necessary, he took the interpretability (or, more exactly, constructibility) of formal results to be a crucial requirement. He believed that it must be the case, for practical reasons, that formal mathematics can be interpreted over the domain of actual numbers. In this, Hankel closely followed Peacock's view. Indeed, more traces of the latter can be identified in Hankel's writings, including statements that endorse Peacock's rejection of generalization: 

\begin{quote} \singlespacing

The purely formal mathematics, whose principles we have stated here, does not consist in a generalization of the usual arithmetic; it is a completely new science, the rules of which
are not proved, but only exemplified, insofar as the formal operations, applied to actual numbers, give the same results as the intuitive operations of common arithmetic. In the latter the definitions of the operations determine their rules, in the former the rules [determine] the meaning of the operations, or to put it another way, they give the instruction for their interpretation and their use. (\textit{op. cit.}, 12)
 
\end{quote}

\noindent Hankel also embraced Peacock's novel metasemantics. One way to understand Hankel's claim that the rules of formal mathematics determine the meaning of its operations, in the sense that they give instructions for their interpretation and application, is by taking the formal rules to provide instructions for the interpretation of formal operations over the domain of actual numbers, where these instructions include a condition of numerical identity. More specifically, formal operations are meaningful only if the actual results obtained by interpreting the results of formal mathematics over the domain of common arithmetic are numerically identical with the actual results derivable by the actual operations of arithmetic. This view is similar to that expressed by Peacock. The interpretability of formal results is necessary for applications, and sufficient to render the formal operations meaningful, provided that the numerical identity condition is generally satisfied.

But Hankel went, in fact, further than Peacock. He considered HP not only as a guide or a merely heuristic or methodological principle, and not only as a metatheoretical principle, characterizing the interpretability relations between theories, either. Rather, Hankel came to think that it was a ``metaphysical'' principle (\textit{op. cit.}, 12), a claim that he attempted to justify by pointing to non-arithmetical interpretations of formal mathematics, e.g., geometrical and physical interpretations. Indeed, Hankel took mechanics to be a theory of actual relations between physical quantities that is, just like arithmetic, merely subordinate to his pure doctrine of forms. And he took HP to stipulate that the relations between physical quantities – the physical laws – must be as much as possible preserved in passing from mechanics to the pure doctrine of forms. The operations of the latter are meaningful only if the actual results derivable by means of the operations of mechanics are the same as the actual results obtained by interpreting formal results over the domain of physical quantities. On this ground, he criticized Peacock's conception of the PF as ``too narrow'' (\textit{op. cit.}, 15). Investigations in the natural science, Peacock had maintained, proceed in two directions: from principles to results, but also towards deeper principles, in a series that terminates only in the ``mystery of the first cause'' (Peacock 1833, 186). Since he thought that the first cause could not be understood as a set of ultimate natural facts, he believed that the deepest principles of the natural sciences could not be conceived of in relation to any formal principles. According to Hankel's view, by contrast, mechanics, just like arithmetic, is another necessarily possible interpretation of the pure theory of forms. Without this possibility, formal operations could have no physical meaning. As we will see presently, this view appears to have had an influence on Bohr's conception of the relation between classical and quantum physics.

\section{The Development of Bohr's Correspondence Principle}

The main point of this paper is to draw attention to a seemingly overlooked connection between HP and Bohr's CP. Having looked at the historical and theoretical context of the former, I now want to revisit the latter. In particular, without attempting to provide an exhaustive analysis, I will recall the emergence of CP in the old quantum theory and some relevant changes that occurred in Bohr's thinking until the application of CP in his approach to quantum
mechanics.\footnote{For more comprehensive accounts of CP, see e.g. Darrigol 1997, Tanona 2002, 2004, Bokulich and Bokulich 2005, Bokulich 2008, Jähnert 2019, and Perovic 2021. For a succinct account, see Bokulich 2020.} But I should note that it is not clear when Bohr actually became aware of HP: it is possible that he knew about it from the very beginning of his articulation of CP, but it is also possible that he found out about HP only later, when as we will see he shared it with his students.\footnote{I have been so far unable to search through the Bohr archive in Copenhagen. Doing so would obviously be important for the line of interpretation of CP pursued in this paper. One very plausible source on HP could have been Bohr’s brother, Harald, who would have been aware of this principle and its importance in the history of mathematics.} It is quite remarkable, nevertheless, that the development of CP in Bohr’s thinking matches rather accurately the views about PF and HP that we have seen Peacock and Hankel to have, respectively, developed. In any case, let me stress here already the expected benefits of my point, in order to properly motivate this inquiry. If CP is understood as grounded in HP, then one can explain several aspects of Bohr's thinking, including his claim that QM is a ``rational generalization'' of classical physics. As we will see, Bohr's notion of rational generalization is based on Hankel’s notion of generalization, which was essentially based on
Peacock’s concept of suggestion. Furthermore, if CP is understood as grounded in HP, then one can also explain a crucial element of Bohr's approach to QM, i.e., his view that the meaning of QM is determined by its rules, a view rather tersely expressed in his reply to the EPR paper. As a bonus, we will also be able to make sense of a seemingly strange criticism directed by Feyerabend and Bohm against this approach, and to show that Howard's reconstruction of Bohr's doctrine of the necessity of classical concepts can avoid any conflict with CP.

Bohr's early use of CP was related to his analysis of radiation into harmonic components. The analysis was concerned with the description of the so-called quantum jumps, i.e., the kind of transitions an electron undergoes between stationary states, which unlike its motion in a particular stationary state, could not be accounted for by classical electrodynamics. The radiation emitted during such transitions allows, according to Bohr's analysis, values of the frequencies in the harmonic components different from the classical values. But he noted an approximate agreement between quantum and classical frequency values and stipulated such an agreement between transition probabilities and the amplitudes of the harmonic components of the classical motion, in the limit of large quantum numbers. He also stipulated an agreement between transition probabilities and the amplitudes of the harmonic components, in the case of small quantum numbers. Bohr appears to have included all these possible relations in his early notion of correspondence:

\begin{quote} \singlespacing

This correspondence between frequencies determined by the two methods must have a deeper significance and we are led to anticipate that it will also apply to the intensities. This is equivalent to the statement that, when the quantum numbers are large, the relative probability of a particular transition is connected in a simple manner with the amplitude of the corresponding harmonic component in the motion. This peculiar relation suggests a general law for the occurrence of transitions between stationary states. Thus we shall assume that even when the quantum numbers are small the possibility of transition between two stationary states is connected with the presence of a certain harmonic component in the motion of the system. (Bohr 1920, 27–28)
    
\end{quote}

\noindent As Bohr emphasized here clearly enough, however, the correspondence between transition probabilities and the amplitudes of harmonic components, when quantum numbers are large, is anticipated on the basis of the correspondence between frequencies, and should be preserved even when the quantum numbers are small. Note that, as a \textit{general law}, one that is valid for \textit{all} quantum numbers, this correspondence is not deduced from, but is said to be \textit{suggested} by the correspondence that holds for large quantum numbers. Thus, even though CP clearly emerged in Bohr's thinking about radiation and his atomic model, as ``a result of gradual bottom-up hypothesis-building from the experimental context within the confines of the model'' (Perovic 2021, 89), and even though it was concerned primarily with the relation between physical quantities, it seems fair to say that here it was also assumed as a guide in the development of Bohr's radiation theory, as a heuristic principle instrumental for his generalization to the case of all quantum numbers. Furthermore, the stipulated preservation of the relation between transition probabilities and the amplitudes of harmonic components underscores a relation between the classical theory of radiation and Bohr's own radiation theory, a metatheoretical relation that he denoted as a ``formal analogy'', despite his suspicion that the idea of correspondence as formal analogy ``might cause misunderstanding''. Indeed, it's difficult to see how a formal analogy between the classical and the quantum theory could be justified, if one took CP to concern exclusively a relation between physical quantities. But what he meant, I think, is that it is the stipulated preservation of this relation that justifies the formal analogy. Of course, it is a further question what else and precisely how much of the classical theory can be preserved in the transition to quantum theory, and I will return to this question below.

Now, if CP is also understood as stipulating the preservation of certain relations between physical quantities, and if it justifies Bohr's claim of formal analogy, then it further justifies a conception of the quantum theory as a certain kind of generalization of the classical theory. Bohr, himself, implied as much at the third Solvay Congress in 1921, when he noted

\begin{quote} \singlespacing

on the one hand, the radical departure of the quantum theory from our ordinary ideas of mechanics and electrodynamics as well as, on the other hand, the formal analogy with these ideas. ... [T]he analogy is of such a type that in a certain respect we are entitled in the quantum theory to see an attempt of a natural generalisation of the classical theory of electromagnetism. (quoted in Bokulich and Bokulich 2005, 348)
\end{quote}

\noindent What did Bohr mean here by a ``natural generalization'' of a classical theory? And did he mean the same thing when he later referred to quantum mechanics (QM) as a  ``rational generalization'' of classical physics? After QM received a coherent mathematical formulation, did Bohr consider this as a generalization justified by a formal analogy between classical physics and QM, in turn justified by CP, understood to require the preservation of correspondences between physical quantities? An answer to such questions would require a detailed analysis of Bohr's writings after 1925, and especially of the development of his conception of CP, which I need to defer to a subsequent paper. But I should, nevertheless, note that the view that there is a correspondence in the sense of formal analogy between classical mechanics and QM was expressed by Dirac:

\begin{quote} \singlespacing
The correspondence between the quantum and classical theories lies not so much in the limiting agreement when $h \rightarrow $ 0 as in the fact that the mathematical operations on the two theories obey in many cases the same laws. (Dirac 1925, 649)\footnote{Later, Dirac attempted to rigorously justify this view of the correspondence between classical mechanics and Heisenberg's QM by setting up a general theory of functions of non-commuting variables (see Dirac 1945).}

\end{quote}

\noindent Dirac appears to have justified the formal analogy via the preservation of laws (many, though not all of them) rather than, like Bohr, via the preservation of correspondences between physical quantities. Despite differences in their views, such as they were, Bohr also came to characterize the metatheoretical relation between classical mechanics and QM in terms of preservation of rules (many, though not all of them):

\begin{quote}
\singlespacing

In this formalism, the canonical equations of classical mechanics ... are maintained unaltered, and the quantum of action is only introduced in the so-called commutation rules... for any pair of canonically conjugate variables. While in this way the whole scheme reduces to classical mechanics in the case h = 0, all the exigencies of the correspondence argument are fulfilled also in the general case... (Bohr 1939, 14)

\end{quote}

\noindent Bohr clearly indicates that this characterization is in accordance with his CP, which suggests that, as recent commentators aptly put it, ``Bohr is not simply saying that the quantum theory should `go over' to the classical theory in the appropriate limit. Rather, he is maintaining that that quantum mechanics should be a theory that \textit{departs as little as possible from} classical mechanics.'' (Bokulich and Bokulich 2005, 349, emphasis added) The suggestion is that CP, and more precisely the inherent notion of generalization, should be understood as the requirement that QM should preserve \textit{as many classical rules as possible}. It is this very notion of generalization, then, that Bohr thought characterized the relation between classical physics and QM. Obviously, this does not imply that classical physics is just a particular case of QM, which \textit{would} be implied if QM were a mere or properly called generalization of classical physics. 

Having focused on Bohr's notion of rational generalization, I now want to suggest further that this is exactly the notion of generalization that Peacock and Hankel had thought characterized the relation between arithmetic, on the one hand, and symbolic algebra or formal mathematics, on the other hand. If this is true, then it is only in part correct to say that ``Through his rational generalization thesis, Bohr is offering us a new way of viewing the relationship between classical and quantum mechanics.'' (Bokulich and Bokulich 2005, 354) Bohr's notion of a natural or rational generalization was new only in the sense that it had not been applied, before him, to the relation between classical and quantum physics. Although evidence for the connection between Bohr's view and that of Hankel will be presented only in the next section, let me follow up a bit on my suggestion.

Recall that Peacock rejected the view that symbolic algebra is a generalization of arithmetical algebra, for he thought this would imply that the principles of symbolic algebra could then be deduced from, rather than just suggested by the principles of arithmetical algebra. He also thought that the relation of suggestion between these theories implied that an arithmetical interpretation of symbolic algebraic results is not necessary, but only necessarily possible. Later, Hankel held a similar view and maintained that the laws of formal mathematics cannot be proved by the rules of arithmetic, but must nevertheless be interpretable in arithmetical language, in the sense that formal results must, when arithmetically interpreted, be numerically identical with the results of arithmetic. This shows that, on both of these 19th century views, their notion of generalization implied a certain interpretability relation between the theories they were concerned with. In order to test the viability of my suggestion to understand Bohr's notion of rational generalization in the same vein, even before I adduce any evidence for the connection between his view and that of Hankel, one should ask whether Bohr's notion also implies a certain interpretability relation between QM and classical physics. But the fact that it actually does so is, of course, well known. A crucial element of Bohr's approach to QM, which will be discussed further below, is the requirement that any description of experimental results must be ``essentially equivalent'' to their classical description. This requirement was taken to have metasemantic implications: the necessarily classical description of experimental results settled, on Bohr's view, the question about the meaning of QM, which became puzzling in the context of the measurement problem. As he put it in his reply to the EPR paper, ``there can be no question of any unambiguous interpretation of the symbols of quantum mechanics other than that embodied in the well-known rules which allow to predict the results to be obtained by a given experimental arrangement described in a totally classical way.'' (Bohr 1935, 701) This sounds a lot like Peacock and Hankel: rules determine the meaning of symbols, rather than the other way around. However, for the rules of QM to determine, or ``embody'', its meaning, Bohr demanded that experimental results \textit{must be} classically interpreted. Here he deviated from Peacock and Hankel in a significant way, and this exposed him to criticism from Feyerabend and Bohm, as we will see. But before that, let me turn to the imminent question about textual evidence. 

\section{Correspondence as Permanence of Rules}

Is there any textual evidence that Bohr understood CP as an expression of HP? If there is, then it has remained by and large unnoticed by Bohr scholars, so far as I have been able to determine. Max Jammer, who seems to have been the first commentator to read CP as a metatheoretical principle, though not also as a quantum law, mentioned HP in one of his ``digressions'' from a streamlined presentation of the conceptual development of QM:

\begin{quote} \singlespacing
    
[M]atrices, multidimensional vectors, and quaternions are extensions of the concept of real numbers. Beyond the domain of complex numbers, however, extensions are possible only at the expense of Hankel's
principle of permanence, according to which \textit{generalized entities should satisfy the rules of calculation pertaining to the original mathematical entities from which they have been abstracted}. Thus, while associativity and distributivity could be preserved, commutativity had to be sacrificed. It was the price which had to be paid to obtain the appropriate mathematical apparatus for the description of atomic states. (Jammer 1966, 217, emphasis added)

\end{quote}

\noindent Jammer implied that the mathematical description of atomic states in QM, just like the extension of mathematics beyond the complexes, was possible only at the expense of HP, which suggests that he thought HP was invalidated by the development of QM. But this would assume that it is an universally valid principle, a view that we have seen both Peacock and Hankel had rejected. At the same time, Jammer thought that ``there was rarely in the history of physics a comprehensive
theory which owed so much to one principle as quantum mechanics owed to Bohr's correspondence principle.'' (\textit{op. cit.}, 118) A proper understanding of CP would show that there is a ``logical rupture'' between classical mechanics and QM, which he described in the following terms:

\begin{quote}
    \singlespacing
[T]he correspondence principle, while leading to numerical agreements between
quantum mechanical and classical deductions, affirmed no longer a conceptual convergence of the results but established merely a formal, symbolic analogy between conclusions derived within the context of two disparate and mutually irreducible theories. It only showed that under certain conditions (for instance, for high quantum numbers or, in classical terms, for great distances from the nucleus) the formal treatments in both theories converge to notationally identical expressions (and numerically equal results) even though the symbols, corresponding to each other, differ strikingly in their conceptual contents. (\textit{op. cit.}, 227)\footnote{Others appear to have followed Jammer in making a similar point, that CP could not and was not meant to close the conceptual gap between classical physics and QM. To give just two examples:  Olivier Darrigol claimed that ``permanent formal schemes allow transfers of knowledge between successive theories even if their basic concepts appear to be incommensurable.'' (Darrigol 1986, 198sq) Also, Hans Radder wrote: ``Generally speaking, intertheoretical correspondence is primarily of a formal-mathematical and empirical but not of a conceptual nature.'' (Radder 1991, 195)}

\end{quote}

\noindent A logical connection or harmony (or whatever the opposite of logical rupture might be) between QM and classical mechanics would establish, Jammer believed, a conceptual convergence of their experimental results, by which he meant an identity of conceptual contents, rather than the numerical identity of results. He took CP to imply that only the latter must obtain between the two theories, while presumably thinking that HP would require the former. However, as we have seen above, Hankel had rejected this understanding of HP, when (following Peacock's view on symbolic algebra) he denied that formal mathematics is a generalization of universal arithmetic. Recall that, according to the reading I proposed, Hankel emphasized that HP implies only that formal mathematics must be interpretable in the language of arithmetic, and when thus interpreted, all formal results should be numerically identical with results derived in arithmetic. What Jammer wrote about CP is correct then, provided that one takes CP as a version of HP properly understood. 

The fact that Bohr did take CP as a version of HP has been reported by Paul Feyerabend, who reminisced that, some time between 1949--1952, in some of his seminars,

\begin{quote}
\singlespacing

Bohr ... talked about the discovery that the square root of two cannot be an integer or a fraction. To him this seemed an important event, and he kept returning to it. As he saw it, the event led to \textit{an extension of the concept of a number that retained some properties of integers and fractions and changed others}. Hankel, whom Bohr mentioned, had called the idea behind such an extension the principle of the permanence of rules of calculation. \textit{The transition from classical mechanics to quantum mechanics, said Bohr, was carried out in accordance with precisely this principle.} That much I understood. The rest was beyond me. (Feyerabend 1995, 76-78, emphasis added)

\end{quote}

\noindent Note that, according to Feyerabend, Bohr's reading of HP did not assume the universal validity of the principle. In accordance with Hankel's own formulation, Bohr knew well that when domains are extended in mathematics, the rules of calculation are always preserved as far as this is possible. He also explicitly emphasized the significance of HP for the development of QM as he saw it: it was the very transition from classical mechanics that he saw carried out in accordance with HP, a significance that he typically attributed, in print, to CP. It is in fact utterly remarkable that, in his published works, Bohr always maintained that this transition was carried out in accordance with CP, and as far as I have been able to determine, he never mentioned HP or Hankel at all. In any case, as Feyerabend's reminiscence indicates, Bohr also knew too well that, as in mathematics, classical rules are preserved in QM as far as this is possible. The same point is clearly made in his 1938 Warsaw conference talk, already quoted above:

\begin{quote} \singlespacing

In the search for the formulation of such a generalization [of the customary classical description of phenomena], \textit{our only guide} has just been the so-called correspondence argument, which gives expression for \textit{the exigency of upholding the use of classical concepts to the largest possible extent compatible with the quantum postulates}. (Bohr 1939, 13; emphasis added)
    
\end{quote}

\noindent After one seminar meeting, Feyerabend confessed his lack of understanding to Bohr's assistant, Aage Petersen. In a decade or so, Feyerabend returned to that conversation:

\begin{quote}
\singlespacing    

As Aage Petersen has pointed out to me, Bohr’s ideas may be compared with Hankel’s principle of the permanence of rules of calculation in new domains...  According to Hankel’s principle \textit{the transition from a domain of mathematical entities to a more embracing domain should be carried out in such a manner that as many rules of calculation as possible are taken over from the old domain to the new one}. For example, the transition from natural numbers to rational numbers should be carried out in such a manner as to leave unchanged as many rules of calculation as possible.  In the case of mathematics, this principle has very fruitful applications. (Feyerabend 1962, 120)

\end{quote}

\noindent What Bohr had surely explained to his students, Feyerabend now finally understood correctly: HP should not be taken to hold universally. It stipulates, just like Peacock and Hankel emphasized, that rules or laws are to be preserved to the largest extent possible. Thus, if applied to QM under the guise of CP, as I take it to have been the case, then HP allows that those classical laws that are essentially classical, like the commutativity of operations, can be given up. 

As already noted in the previous section, this evidence, based on Feyerabend's recollections of Bohr's lectures and Petersen's explanations, does not tell us precisely when Bohr actually became aware of the connection between HP and CP: it is possible that he knew about it from the very beginning of his articulation of CP, in his theory of radiation, but it is also possible that he found out about HP only later, maybe after 1925, or even later than that in the late 1930s or early 1940s. But I find it quite remarkable that Bohr's characterizations of CP as a methodological principle or a guide -- the ``only guide'', as he specified in Warsaw in 1938 -- as well as as a metatheoretical principle concerning the ``rational generalization'' of classical physics in QM, match rather accurately the views about PF and HP that we have seen Peacock and Hankel to have,  respectively, developed in 19th century mathematics. The evidence supports at least the claim that Bohr's notion of rational generalization was grounded in Hankel's notion of generalization, which in turn was grounded, as we have seen, in Peacock's notion of suggestion. This clarifies, I hope, what so far has been a rather enigmatic detail in Bohr's works. 

In fact, more can be explained on the basis of my interpretation of the connection between HP and CP. For Feyerabend had, of course, a lot more to say about Bohr's approach to QM. One particular weakness with this approach that he immediately identified was described as follows: 

\begin{quote}
\singlespacing    

A complete replacement of the classical formalism seems therefore to be unnecessary. All that is needed is a modification of that formalism which retains the laws that have found to be valid and makes room for those new laws which express the specific behavior of the quantum mechanical entities. ... [The new laws] must allow for the description of any conceivable experiment in classical terms -- for it is, in classical terms that results of measurement and experimentation are expressed; ... [this requirement] is needed \textit{if we want to retain the idea... that experience must be described in classical terms}. (Feyerabend 1962, 120; emphasis added)

\end{quote}

\noindent As Feyerabend correctly observed, Bohr insisted that the classical description of experimental results, i.e., presumably their interpretation in the language of classical physics, is necessary. But Feyerabend rejected this necessity claim, which might seem rather strange. His criticism, further elaborated in the same paper, emphasizes the point that in principle a different language could be developed at least as adequate for the description of experimental results as the language of classical physics. Feyerabend's point, that a classical description is not necessary, allows however that this may be necessarily possible. In light of the view articulated by Peacock and Hankel, Feyerabend's criticism appears to be justified. It clearly emphasizes that Bohr's view deviated from that of Peacock and Hankel, for whom interpretability, though no particular interpretation, was necessary. But it is this deviation precisely that exposed Bohr to Feyerabend's criticism.\footnote{To be sure, Feyerabend's opinions about Bohr's approach to QM have also evolved over time (see Kuby 2021).} This very criticism was later pressed by Bohm as well: ``What is called for, in my view, is therefore a movement in which physicists freely explore novel forms of language, which take into account Bohr’s very significant insights but which do not remain fixed statically to Bohr's adherence to the need for classical language forms.'' (Bohm 1985, 159; quoted in Bokulich and Bokulich 2005, 368). That the description of experimental results must be given in a classical language, Bohm might have added, just because they are in practice presented in this language is not merely a static fixation, but a downright fallacy.

Having offered some evidence that Bohr understood CP as a version of HP, or at least that he (and some of his assistants and, eventually, Feyerabend) thought that a comparison of the former with the latter would be fruitful for understanding the transition from classical physics to QM, and having also admitted that this could only justify the necessary possibility of interpreting experimental results in classical terms, rather than Bohr's insistence on its necessity, I want to turn to the question of what exactly Bohr meant by ``classical''. Some commentators think that he took such concepts to be simply concepts of classical mechanics and electrodynamics (cf. Bokulich and Bokulich 2005, 351), but others maintain that, for Bohr, a classical description meant ``a description in terms of what physicists call `mixtures''' (Howard 1994, 203). I want to argue that, despite appearances, if the connection between CP and HP is taken seriously, then one can rather nicely accommodate the latter view.

\section{Howard on Bohr’s Essential Equivalence}

In his reconstruction of Bohr's philosophy of physics, Don Howard emphasized that the doctrine of the necessity of classical terms was upheld by Bohr in an attempt to overcome a problem for objectivity that arises in QM. The problem is that what is generally considered a necessary condition for objectivity -- the metaphysical independence of observer and observed reality, and more precisely their separability, which Einstein thought was indispensable to the very formulation and testing of physical laws -- cannot be preserved when passing from classical physics to QM. As Howard presented it, Bohr's doctrine was meant as a purported solution to this problem. Classical terms are necessary because they ``embody'' the separability condition, which despite being false in QM allows for an unambiguous communicability of experimental results.\footnote{Cf. Howard 1994, 207. Note that separability is understood as state decomposability. To say that classical concepts ``embody'' the separability condition is taken to mean that separability is mathematically equivalent to Bohr's doctrine of the necessity of classical concepts (see Landsman 2006 for a proof of this equivalence). This entails that separability and entanglement are incompatible. As Howard emphasized, this is precisely the reason Bohr's solution to the problem of objectivity is unacceptable. But note also that separability, as Einstein himself appears to have conceived of it, may be a weaker condition than state decomposability and, thus, compatible with entanglement (see Murgueitio Ram\'irez 2020 for an argument to this effect).}

Further, Howard distinguished two ways of understanding Bohr's doctrine: one of them ``leaves open the possibility that, as our language develops, we might outgrow this dependence'' on classical concepts; the other, which is considered preferable, takes ``the necessity of classical concepts to be an enduring one, not to be overcome at a later stage in the evolution of language.'' (Howard 1994, 209) However, barring the potential fallacy mentioned in the previous section, it is not clear why the latter view should be preferred. What exactly might explain Bohr's insistence on the enduring character of classical language? What reasons did he have, and what reasons might anyone have, for excluding the possibility that other languages, as both Feyerabend and Bohm suggested, could at least in principle be developed to communicate quantum-mechanical results unambiguously and at least as adequately as the language of classical physics? If CP is understood as grounded in HP, as I have suggested, then it is the interpretability, rather than any particular interpretation, of experimental results that should be required by Bohr's doctrine. This would leave open the possibility envisaged by Feyerabend and Bohm.\footnote{It is of course also possible that Bohr had other unstated reasons, unrelated to CP's grounding in HP, that he took to justify his doctrine (see Faye 2017).} 

More importantly, however, Howard argued that, in demanding a classical description of experimental results, Bohr's doctrine does not require that a measuring instrument must be described \textit{entirely} in classical terms. Rather, only some of its properties are to be described classically, i.e., those that are correlated with the properties of the quantum system undergoing measurement (\textit{op. cit.}, 216). Howard took this to imply that what Bohr meant by a classical description should be most plausibly reconstructed as a description in terms of mixtures, rather than pure states; mixtures that must always be appropriate to a given experimental context. The reason for this is that, unlike pure states, mixtures are considered to ``embody'' the separability condition, in the sense that they allow the separability of measuring instrument and measured object with regard to exactly those properties of the object one is looking to determine in a particular measurement.

What is the role of CP on this reconstruction of Bohr's doctrine? As Howard noted, this doctrine requires an ``essential equivalence'', i.e., an equivalence between, on the one hand, the QM description of the properties of the measuring instrument that are correlated with the measured properties of the system undergoing measurement and, on the other hand, the classical description of those properties of the measuring instrument. Indeed, the main goal of Bohr's 1938 Warsaw conference paper was to discuss ``certain novel epistemological aspects'' involved in what he called ``the observation problem'' and, more specifically, certain aspects regarding ``the analysis and synthesis of physical experience.'' (Bohr 1939, 19) What were these aspects? The main outcome of the analysis was an emphasis on the necessity of taking the whole experimental arrangement, i.e., measured object plus measuring instrument, into consideration. Without this, said Bohr, no unambiguous meaning could be given to the QM formalism (\textit{op. cit.}, 20). The outcome of the synthesis was presented as follows:

\begin{quote} \singlespacing

In the system to which the quantum mechanical formalism is applied, it is of course possible to
include any intermediate auxiliary agency employed in the measuring process. Since, however, all those properties of such agencies which, according to the aim of the measurement, have to be
compared with the corresponding properties of the object, must be described on classical lines, their quantum mechanical treatment will for this purpose be essentially equivalent with a classical description. (\textit{op. cit.}, 23sq)

\end{quote}

\noindent Thus, Bohr's insight was that giving a classical description of experimental results can only mean establishing the essential equivalence of the two descriptions of the relevant subset of properties of the measuring instrument. But establishing such an equivalence, Howard maintained, is at odds with Bohr's CP:

\begin{quote}
\singlespacing

[W]hat kind of ``classical'' description could be ... ``essentially equivalent'' to a quantum mechanical description. In the sense intended by the \textit{correspondence principle}, quantum mechanics might agree with Newtonian mechanics or with Maxwell's electrodynamics in the limit of large quantum numbers, but that is not an ``essential'' equivalence. Moreover, the kind of convergence between quantum and classical descriptions demanded by the correspondence principle is a wholesale convergence, not an equivalence between selected sets of properties. ...

How can a classical description be `essentially equivalent' to a quantum mechanical one? Bohr's correspondence principle is what first comes to mind, but it cannot provide the answer, for two reasons. First, the correspondence principle asserts that quantum and classical descriptions
agree in the limit of large quantum numbers, that, is, in phenomena where the quantum of action is negligible. ...

Second, what the correspondence principle says about the relationship between classical and quantum descriptions is that they give \textit{approximately} the same predictions in the limit of large quantum numbers. But approximate agreement is hardly essential equivalence. The appropriate mixtures model gives a quite different answer. A quantum mechanical description, in terms of a pure case, and a `classical' description, in terms of
an appropriate mixture, give \textit{exactly} the same predictions for those observables measurable in the context that determines the appropriate mixture.\footnote{Cf. Howard 1994, 217-225. See also Howard 2021, 166 for a more recent iteration of this view.}

\end{quote}

\noindent As we have seen above, there is evidence (and an apparent consensus today) that CP should be read as asserting not merely an approximate agreement that holds in the limit of large quantum numbers, but an agreement that also holds more generally, for small quantum numbers as well. But I think that this poses no problem for Howard's reconstruction of Bohr's doctrine of classical concepts. This is because my account of CP as grounded in HP entails that there is no conflict at all between CP and Bohr's demand of an essential equivalence. Quite the opposite, this account can nicely accommodate the fact that a ``wholesale convergence'', i.e., an equivalence between the QM description of \textit{all} properties of the measuring instrument and their classical description, cannot be established, and that an essential equivalence, as reconstructed by Howard, is necessary for the classical \textit{interpretability} of experimental results. This point can be succinctly clarified by appeal to Klaas Landsman's Bohrification strategy.\footnote{Cf. Landsman 2017. Howard's own formal explication of what it means for a classical description to be essentially equivalent to a QM description is based on his 1979 theorem concerning context-dependent mixtures. For details, see Howard 2021, 162-170.} Quantum measurement results, considered as physically significant aspects of a noncommutative algebra of observables (NAO), are accessible only if they can be described classically, i.e., only if they can be considered as aspects of a commutative algebra (CA). But NAO should be considered as a rational generalization of CA, in Bohr's sense. This kind of generalization requires exactly the essential equivalence that Bohr demanded, which can be established if and only if the elements of CA are a proper subset of the elements of NAO -- the subset determined by the particular experimental context. A wholesale convergence, which would require that CA and NAO be coextensive, is mathematically impossible.

\section{Conclusion}

The significance of the principle of permanence for our understanding of the correspondence principle is, I believe, undeniable. As interpreted by Bohr, QM turns out to be a rational generalization of classical physics in precisely the sense of generalization that had been articulated already by Peacock and Hankel in the 19th century. Bohr's slight deviation from Peacock's and Hankel's views exposed him to Feyerabend's and Bohm's criticism, which becomes intelligible enough when considered against the historical background presented in this paper. My analysis, I hope, goes some of the way towards placing ``Bohr's views on the role of classical concepts ... in their proper historical context, especially as regards the relevant philosophical context.'' (Howard 1994, 227) As always, though, this is merely part of the whole story, more details of which await to be uncovered.

\section*{Acknowledgements}

Thanks to Anna Bellomo, Richard Lawrence, and Georg Schiemer, for discussion of Peacock and especially Hankel, as well as to Sebastian Horvat and the two referees for this journal, whose detailed comments on Bohr helped me improve the paper significantly.

\section*{References}

Bohm, D. (1985) ``On Bohr's views concerning the quantum theory'', in A. French and P. Kennedy (eds.) \textit{Niels Bohr: A Centenary Volume},  Harvard University Press, 153--159.

\smallskip

Bohr, N. (1920) ``On the series spectra of the elements'', Lecture before the German Physical Society in Berlin, 27 April 1920, in J. R. Nielsen (ed.) \textit{Niels Bohr Collected Works, Vol. 3: The Correspondence Principle (1918–1923)} Amsterdam,  North-Holland Publishing, 241--282.

\smallskip

Bohr, N. (1935) ``Can Quantum-Mechanical Description of Physical Reality Be Considered Complete?'', in \textit{Physical Review}, 48, 696--702.

\smallskip

Bohr, N. (1939) ``The causality problem in atomic physics'' (Report drafted and submitted by N. Bohr), in \textit{New theories in physics: Conference organized in collaboration with the International Union of Physics and the Polish Intellectual Co-operation Committee, Warsaw, May 30th-June 3rd, 1938}, 11--45.

\smallskip

Bokulich, A. (2008) \textit{Reexamining the Quantum-Classical Relation. Beyond Reductionism and Puralism}, Cambridge University Press.

\smallskip

Bokulich, A. and P. Bokulich (2020) ``Bohr's Correspondence Principle'', in \textit{The Stanford Encyclopedia of Philosophy}.

\smallskip

Bokulich, P. and A. Bokulich (2005) ``Niels Bohr’s generalization of classical mechanics'', in \textit{Foundations of Physics}, 35, 347--371.

\smallskip

Darrigol, O. (1986) ``The Origin of Quantized Matter Waves'', in \textit{ Historical Studies in the
Physical and Biological Sciences}, 16, 197--253.

\smallskip

Darrigol, O. (1997) ``Classical Concepts in Bohr’s Atomic Theory (1913–1925)'', in \textit{Physis: Rivista Internazionale di Storia della Scienza}, 34, 545--567.

\smallskip

Detlefsen, M. (2005) ``Formalism'', in Shapiro, Stewart (Ed.), \textit{The Oxford Handbook of Philosophy of Mathematics and Logic}, Oxford University Press, 236--317.

\smallskip

Dirac, P.A.M. (1925) ``The Fundamental Equations of Quantum Mechanics'', in \textit{Proceedings of
the Royal Society of London}, Series A, 109, 642--653.

\smallskip

Dirac, P.A.M. (1945) ``On the Analogy Between Classical and Quantum Mechanics", in \textit{Reviews of Modern Physics}, 17, 195--199.

\smallskip

Faye, J. (2017) ``Complementarity and Human Nature'', in 
J. Faye and H. J. Folse (eds.) \textit{Niels Bohr and the Philosophy of Physics: Twenty-First-Century Perspectives}, Bloomsbury, 115--131.

\smallskip

Feyerabend, P. (1962) ``Problems of Microphysics'', in S. Gattei and J. Agassi (eds.) \textit{Paul K. Feyerabend
Physics and Philosophy
Philosophical Papers
Volume 4}, Cambridge University Press, 99--187.

\smallskip

Feyerabend, P. (1995) \textit{Killing Time}, The University of Chicago Press.

\smallskip

Gregory, D. (1840) ``On the real nature of symbolic algebra'', in W. Ewald (ed.) \textit{From Kant to Hilbert: A Source Book in the Foundations of Mathematics}, vol. 1, Oxford University Press, 1996, 323--330.

\smallskip

Hankel, H. (1867) \textit{Theorie der complexen Zahlensysteme}. Leopold Voss, Leipzig.

\smallskip

Hollings, C., U. Martin, and A. Rice (2017) The Lovelace–De Morgan mathematical correspondence: A critical re-appraisal. \textit{Historia Mathematica}, 44, 202--231.

\smallskip

Howard, D. (1994) ``What makes a classical concept classical? Towards a reconstruction of
Niels Bohr’s philosophy of physics'', in J. Faye and H. Folse (eds.) \textit{Niels Bohr and Contemporary Philosophy}, Kluwer
Academic, Dordrecht, 201--229.

\smallskip

Howard, D. (2021) ``Complementarity and decoherence'', in G. Jaeger, D. Simon, A. V. Sergienko, D. Greenberger, \& A. Zeilinger (eds.) \textit{Quantum arrangements: Contributions in honor of Michael Horne}, Springer, 151--175.

\smallskip

J\"ahnert, M. (2019) \textit{Practicing the
Correspondence Principle in
the Old Quantum Theory}, Springer.

\smallskip

Kuby, D. (2021) ``Feyerabend's Reevaluation of Scientific Practice: Quantum Mechanics, Realism and Niels Bohr, in K. Bschir \& J. Shaw (eds.) \textit{Interpreting Feyerabend: Critical Essays}, Cambridge University Press, 132-156.

\smallskip

Lambert, K. (2013) ``A Natural History of Mathematics: George Peacock and the Making of English Algebra'', in \textit{Isis}, 104, 278--302.

\smallskip

Landsman, N. P. (2006) ``When champions meet: Rethinking the Bohr–Einstein debate'', in \textit{Studies in History and Philosophy of
Modern Physics}, 37, 212--242.

\smallskip

Landsman, K. (2017) \textit{Foundations of Quantum Theory. From Classical Concepts to Operator Algebras}, Springer.

\smallskip

Murgueitio Ram\'irez, S. (2020) ``Separating Einstein's separability'', in \textit{Studies in History and Philosophy of Modern Physics}, 72, 138--149.

\smallskip

Peacock, G. (1833) ``Report on the Recent Progress and Present State of Certain Branches of Analysis'', in \textit{Report of the British Association}, 185--352.

\smallskip

Perovic, S. (2021) \textit{From Data to Quanta: Niels Bohr's Vision of Physics}, The University of Chicago Press.

\smallskip

Pycior, H. (1981) ``George Peacock and the British Origins of Symbolic Algebra'', in \textit{Historia Mathematica}, 8, 23--45.

\smallskip

Radder, H. (1991) ``Heuristics and the Generalized Correspondence Principle'', in \textit{The British Journal for the Philosophy of Science}, 42, 195--226.

\smallskip

Tanona, S. (2002) \textit{From correspondence to complementarity: The emergence of Bohr's Copenhagen interpretation of quantum mechanics}, PhD dissertation, Indiana University.

\smallskip

Tanona, S. (2004) ``Idealization and Formalism in Bohr’s Approach to Quantum Theory'', in \textit{Philosophy of Science}, 71, 683--695.

\smallskip

Toader, I. D. (2021) ``Permanence as a Principle of Practice'', in \textit{Historia Mathematica}, 54, 77--94

\smallskip

Toader, I. D. (2023) ``Permanence of Forms as a Transfer Principle'', talk at \textit{From Permanence to Conservativity: Metatheoretic Ideals in Early Formalism}, University of Vienna, December 4, 2023

\smallskip

Whewell, W. (1840) \textit{The Philosophy of the Inductive Sciences, Founded Upon Their History}, London, John W. Parker.

\end{document}